%% file: stephensen_darkner_sporring.tex
\documentclass[10pt,journal,compsoc]{IEEEtran}

\input{preamble.tex}

\title{A Highly Accurate Model Based Registration Method for FIB-SEM Images of Neurons}

\author{Hans~JT~Stephensen,
        Sune~Darkner,
        and~Jon~Sporring
\IEEEcompsocitemizethanks{\IEEEcompsocthanksitem All authors are with the Department
of Computer Science, University of Copenhagen.\protect\\
Article is pending submission to peer-reviewed journal.\protect\\
Contact: sporring@di.ku.dk}
\thanks{Submitted October 2, 2018.}}

\begin{document}

\maketitle



\input{body.tex}

\input{methods.tex}


\bibliographystyle{IEEEtran}
\bibliography{IEEEabrv,references}






\end{document}

%% file: preamble.tex
\usepackage{setspace}
\usepackage{amsfonts}
\usepackage{amssymb}
\usepackage{amsmath}

\usepackage{float}
\usepackage{placeins}
\usepackage{graphicx}

\usepackage[export]{adjustbox}

\newcommand{\argmin}{\operatornamewithlimits{argmin}}

\usepackage[super,comma,sort&compress]{natbib}

\usepackage{hyperref}
\usepackage{verbatim}


%% file: body.tex


\section*{Introduction}

\begin{figure*}
  \centering
  \includegraphics[width=\textwidth]{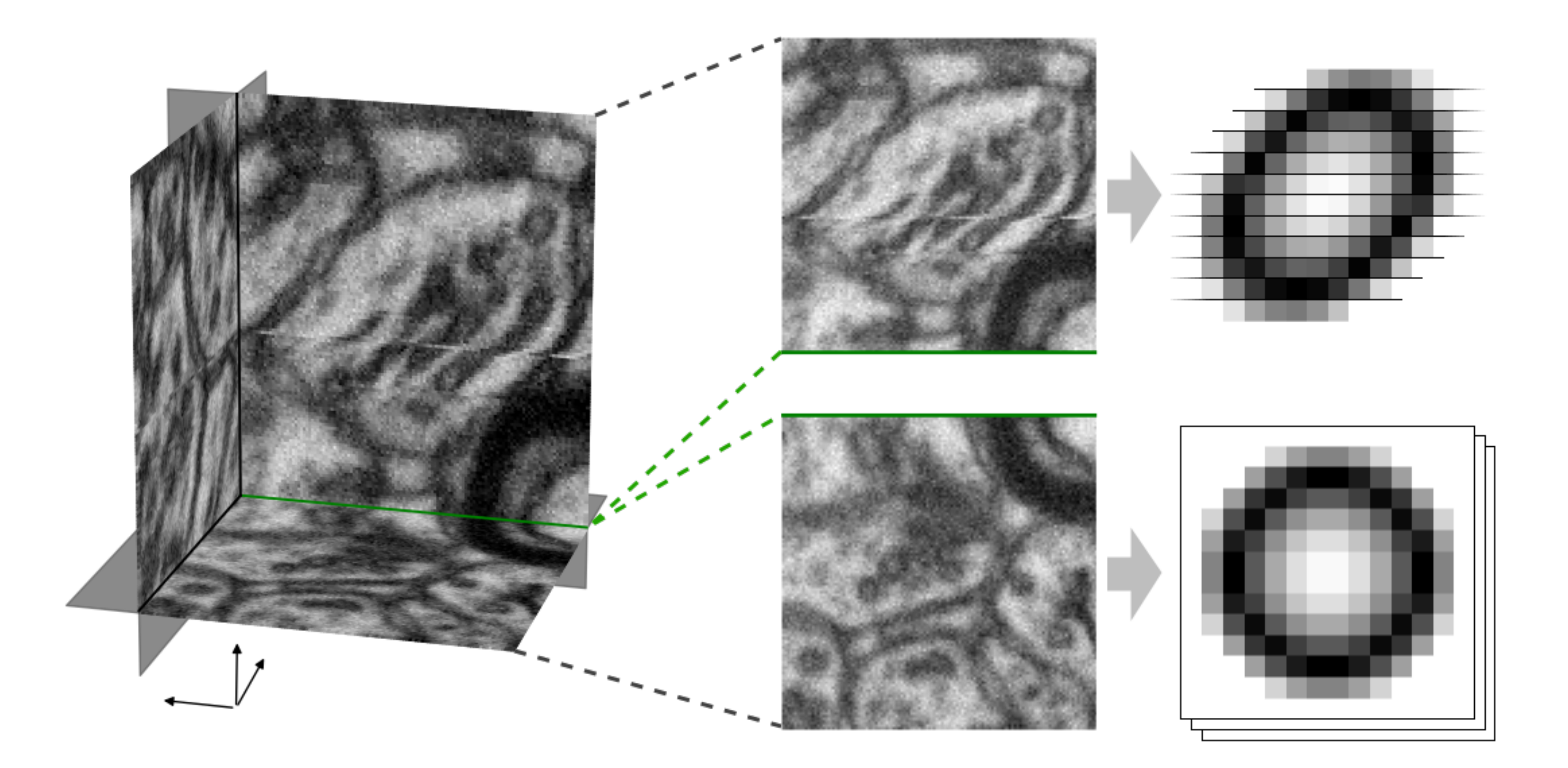}
  \caption{(Left) A look inside a section of a FIB-SEM dataset with significant drift present. (Right-Top) Example of a stretched vesicle when viewing the image across multiple sections. (Right-Bottom) Example of the stretched vesicle when viewing the image in a single section, i.e., in the image plane.}
  \label{fig:fibsem_cube_stretched_vesicles}
\end{figure*}

In many common 3D scanning methods used for imaging biological material, the scanning method itself may at each imaged section induce drift from the previous section causing a misalignment in the section direction. Drift can arise from a variety of practically uncontrollable factors such as bending of the electron beam due to a charge gradient in the material or physical movement of the entire sample or within the sample itself to name a few examples. Correcting this drift is crucial to any subsequent work on the images because the drift skews distance measures. Since any statistics or shape analysis is deeply dependent on the accuracy of such distance measures, this is likely to have significant consequences for the biological conclusions presented on the basis thereof. Fig. \ref{fig:fibsem_cube_stretched_vesicles} shows an example of an ultrastructure brain region from a healthy adult rodent with easily noticeable drift when the dataset is viewed across multiple image planes. Datasets such as this have been and are still used actively (see \cite{wu2017contacts,bosch2015fib,cali2016three,khanmohammadi2015statistical} for a few examples) with no apparent mention of correction for potential drift making it unclear what effect such misalignment may have had on the presented results. In other cases \cite{morales2011espina,merchan2009counting,ender2012quantitative}, correction has been done using ImageJ typically relying on packages such as the stackreg and turboreg packages. These packages support both manual registration, where landmark points are chosen as a basis for the alignment, or using automatic methods in some extended form of standard image registration methods such as pyramidal least-squares minimization of the image intensities \cite{thevenaz1998pyramid}. Other notable approaches include maximizing the mutual information \cite{collignon1995automated} or normalized mutual information \cite{studholme1999overlap}.\\

With drift present in the images, it is important that the image sections are realigned before further study. This is done either manually by specifying drift parameters on the scanning device, or in post-processing either manually by hand or automatically using one of the above mentioned image registration methods. Both of these methods are likely to introduce error or biasing of unknown severity. In neuron tissue specifically, it is common that nearby neurons will have a similar orientation. This affects standard registration methods because they minimize, not only the drift but also any apparent transversal movement of the neurons across image sections. In practical cases, the effect of drift on the estimated transformation is comparable in size to the transversal movement of the neurons causing poor results using standard registration methods meaning they are arguably poorly suited for problems such as this.\\

In this paper, we present a novel model-based approach for biologically accurate translational image registration on FIB-SEM images of biological material containing visible vesicles. The idea is to estimate the vesicle organelle shape and to use deviations from the expected shape to determine the correction needed. Since vesicles are numerous and since on average vesicles are expected to be spherical, we can estimate the drift by enforcing this property on the vesicles. This method solves the aforementioned problem with neuron orientation because the shape of the vesicle is more locally dependent compared to larger structures.

Our method relies on estimating the shape of the vesicles by ellipsoids. We do this by first annotating the vesicle boundary by points of an appropriate number of vesicles. We then estimate the ellipsoid parameters by least squares approach.

Due to the absence of any reliable ground truth drift on real data, we are here limited to a qualitative assessment of drift correction carried out on real data. As a consequence, we augment the assessment by further experiments on synthetically generated images with known added drift.

\section*{Drift Estimation by Standard Image Registration}

Image registration is a means of mapping and transforming one image $I$ wrt. some target image $T$. The usual process consists of formulating the problem as a minimization problem of some functional $\mathcal{F}$, of the form $\mathcal{F} = \mathcal{M}(I, T) + \mathcal{R}$, where $\mathcal{M}$ is a measure formulated to describe the dissimilarity of $I$ and $T$, and $\mathcal{R}$ is a regularization term. Standard measures of $\mathcal{M}$ are formulated as an integral over the image domain $\Omega$ given by $M = \int_\Omega F(\mathbf{x},I(\mathbf{x}),T(\mathbf{x})) \enspace d\mathbf{x} \enspace$, where $F$ are chosen to be a measure such as the Sum of Squared Differences (SSD) or more elaborate scheme involving state if the art measures such as the Mutual Information \cite{collignon1995automated}, Normalized Mutual Information \cite{studholme1999overlap} or using Locally Orderless Registration \cite{darkner2013locally}.\\

\begin{figure}
  \centering
  \includegraphics[width=\columnwidth]{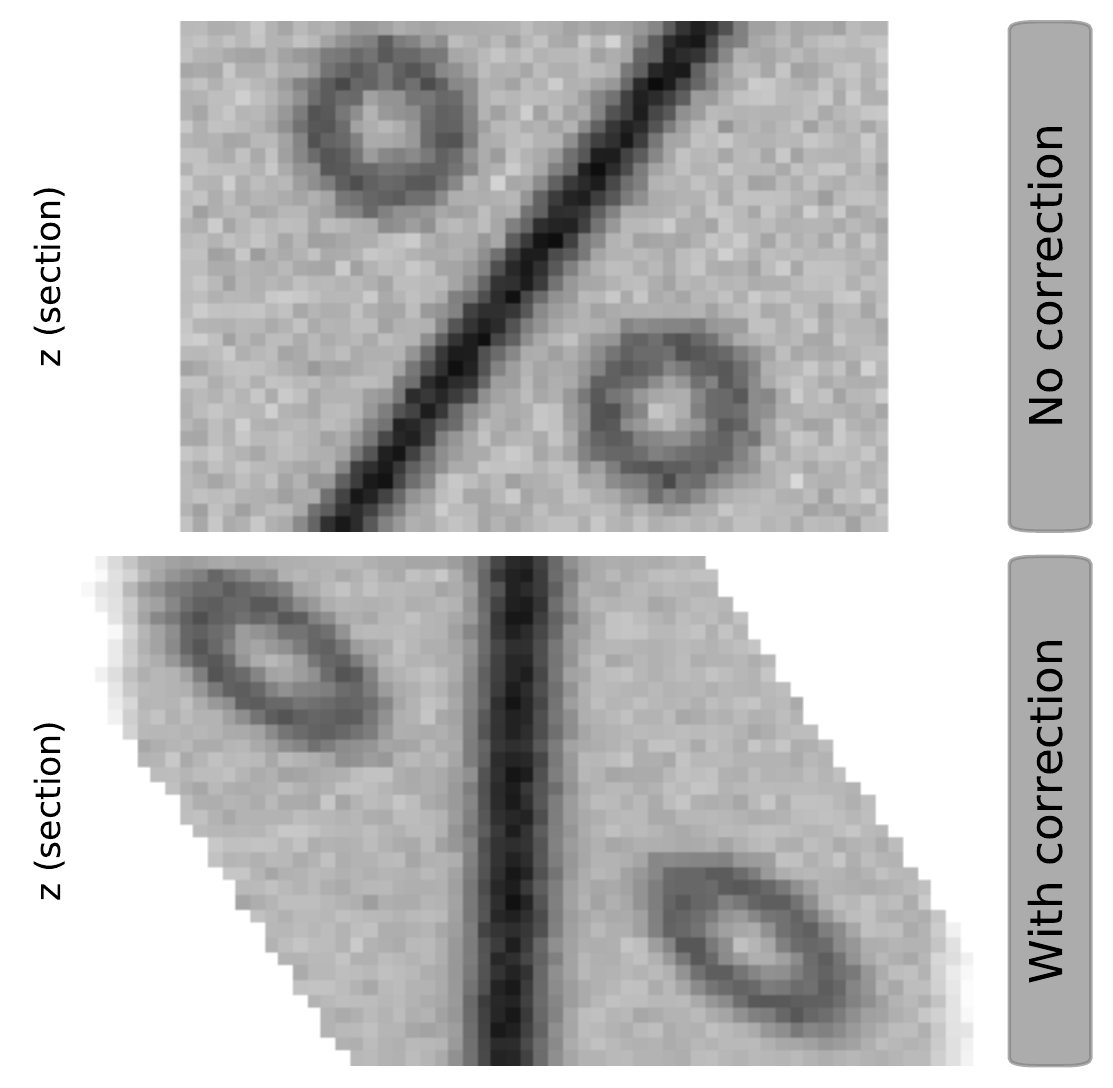}
  \caption{Top: Example of a synthetic image with a plane passing through diagonally to imitate a slated membrane  as well as two hollow spheres playing the role of vesicles. Bottom: The same image as above, but after performing translational correction based on a standard image registration approach implemented in Matlab.}
  \label{fig:bane_of_standard_methods}
\end{figure}

Cellular and sub-cellular structures in FIB-SEM images can spatially appear to move across the image plane when traversing the image plane across multiple sections. Since standard registration methods by default minimize a global measure on the entire image domain, pairwise image registration on consecutive FIB-SEM image sections therefore not only minimize any drift present in the images, but also structures appearing to have a sideways movement in the image plane across sections. An example of the problem can be seen in Fig. \ref{fig:bane_of_standard_methods}, where a simple synthetic 3D FIB-SEM image has been generated with a single slated membrane and two spherical vesicles. Even though no drift is assumed here, a standard registration approach stretches the image to force the membrane to be perpendicular to the image section direction. While a severe case as this is seldom found in real images, it illustrates how registration methods are only secondarily influenced by drift, making standard registration approaches undesirable for a problem such as drift correction.

\section*{Drift estimation from Vesicle Models}
\label{sec:vesicle_model}

The shell of the synaptic vesicle is constituted by a lipid bilayer. The lipid bilayer is commonly modeled physically as an elastic material with a bending energy density functional due to \cite{canham1970minimum,helfrich1973elastic}. This energy functional tells us that to a large degree, the fundamental strive for energy minimization corresponds to a minimization of curvature in the elastic material. In equilibrium conditions, this results in spherical vesicles. Although it is well known that vesicles can take a variety of exotic shapes under special conditions \cite{seifert1991shape,miao1994budding}, the most probable shapes are spheroids, prolate and oblate shapes which are sufficiently well modeled as ellipsoids for the purpose of this work.

%% file: methods.tex



\section*{Methods}
\subsection*{Obtaining Ellipsoids from Boundary Points}
\label{sec:ellip_params_from_boundary_points}

Our first method relies on having obtained boundary points of the vesicles in some or all of the sections of which the vesicle is present. For the present study, we have manually marked these points using a python script. To fit ellipsoids to the point data, we numerically search for an ellipsoid minimizing the sum of squared perpendicular distance from each point to the ellipsoid. To accomplish this, we first initialize a random ellipsoid \(E(\mathbf{c},\mathbf{r},\mathbf{q})\), where \(\mathbf{c}\) is the center point, \(\mathbf{r}\) are the radii along the ellipsoid main axes, and \(\mathbf{q}\) is a quaternion representing the rotation of an initial frame into the ellipsoids main axes. The center point is initialized to the average of the boundary points, radii are uniformly distributed near the expected size of the vesicles, and the direction of the quaternion is uniformly distributed on a sphere, and the rotation uniformly on the interval \([0,2\pi]\). We represent and optimize the rotation as a quaternion in order eliminate bias found in representations such as Euler angles, and to avoid the \textit{Gimbal Lock} in the optimization. We then perform gradient descent optimization to approximately solve the minimization problem as,
\begin{equation}
\label{eq:ssd_functional}
\argmin_{\mathbf{c},\mathbf{r},\mathbf{q}} \left\lbrace \sum_{\mathbf{p}\in \Gamma} d(E(\mathbf{c},\mathbf{r},\mathbf{q}),\mathbf{p})^2 \right\rbrace \enspace ,
\end{equation}
where \(\Gamma\) is the set of boundary points and \(d\) is the perpendicular distance from the ellipsoid to each point \(\mathbf{p} \in \Gamma\). The function \(d\) was here calculated using the Geometric Tools \texttt{C++} Library \cite{GeomTools}. Each set of vesicle points was fitted multiple times using random starting points as described above, keeping only the best fitting ellipsoid according to \eqref{eq:ssd_functional}. We call this the point-model.

\subsection*{Estimating Drift from Ellipsoid Parameters}
\label{sec:ellipsoid_drift_derivation}

Let \(x,y,z\) be the axes of an image with \(x,y\) the plane of each image section and \(z\) the axis in which the image sections are stacked. An ellipsoid centered at the origin can be described implicitly by the quadratic surface equation \(\mathbf{u}^TH\mathbf{u} = 1\), where \(\mathbf{u} = [x,y,z]^T\), and \(H\) is a \(3 \times 3\) symmetric positive definite matrix. We shall name the parameters of \(H\) as
\begin{equation}
\label{eq:ellipsoid_parameter_matrix}
H =
\begin{bmatrix}
A & D & E \\
D & B & F \\
E & F & C
\end{bmatrix} \enspace .
\end{equation}
We will refer to these parameters as the parameters of the ellipsoid defined by \(H\). We note here that the parameters \(E\) and \(F\) determine the shape of the ellipsoid as a function of \(y\) and \(z\) and of \(x\) and \(z\) resp. Setting \(E = F = 0\) forces the ellipsoid to be symmetric across the plane \(z = 0\). Thus, we can understand the value of \(E\) and \(F\) as the ``tilt'' of the ellipsoid as a function of \(z\).

In general, we will assume the drift in the image can be represented as a sideways translation of each image section with respect to the previous section. Denoting \(\delta x, \delta y\) as the amount of translation of some section with respect to the previous and denoting \(\Delta z\) as the distance between subsequent sections, we represent the translation as a shear map with shear coefficients \(s_x = \delta x/\Delta z, \enspace s_y = \delta y/\Delta z\). If we first assume the drift is constant as a function of \(z\), we can then represent the drift as one single mapping \(S\) given by
\begin{equation}
S\mathbf{x} =
\begin{bmatrix}
1 & 0 & s_x \\
0 & 1 & s_y \\
0 & 0 & 1
\end{bmatrix}
\begin{bmatrix}
x \\
y \\
z
\end{bmatrix}
=
\begin{bmatrix}
x + s_xz \\
y + s_yz \\
z
\end{bmatrix}
=
\mathbf{u} \enspace .
\end{equation}
We note that the shear mapping is a non-singular linear transformation. Since \(S^{-1}S\) is the identity transformation, the quadratic equation is still solved when
\begin{equation}
1 = \mathbf{x}^TH\mathbf{x} = \mathbf{x}^T(S^{-1}S)^THS^{-1}S\mathbf{x} = \mathbf{u}^TS^{-T}HS^{-1}\mathbf{u} \enspace .
\end{equation}
Thus, if each point on the ellipsoid is transformed by \(S\), it corresponds to a new quadratic surface defined by the matrix representation \(\hat{H} = S^{-T}HS^{-1}\), or equivalently \(H = S^T\hat{H}S\). Since an ellipsoid is a quadratic surface with a closed surface, and since non-singular linear transformations on closed surfaces cannot produce open surfaces, we conclude that the result is still a closed surface defined by a quadratic surface, i.e., an ellipsoid, spheroid or sphere.\\

Let \(\hat{A},\hat{B},\hat{C},\hat{D},\hat{E},\hat{F}\) be the ellipsoid parameters of \(\hat{H}\), the shear-transformed ellipsoid we have from data. Assuming the values of \(\hat{E}\) and \(\hat{F}\) (the ``tilt'' of the ellipsoid as a function of \(z\)) are solely due to a shear of the ellipsoid, we can solve for a shear map that induced this tilt on the ellipsoid. Thus, we define our ``untilted'' ellipsoid \(H = S^T\hat{H}S\) by setting \(E = F = 0\) and solve for \(s_x\) and \(s_y\). We get
\begin{equation}
s_x = \frac{\hat{D}\hat{F}-\hat{B}\hat{E}}{\hat{A}\hat{B}-\hat{D}^2} \enspace , \quad
s_y = \frac{\hat{D}\hat{E}-\hat{A}\hat{F}}{\hat{A}\hat{B}-\hat{D}^2} \enspace .
\end{equation}
Let \(\textbf{s} = (s_x,s_y)^T\) represent the shear of some ellipsoid. By assumption, each ellipsoid is rotated uniformly at random. Thus, it follows that given no drift in the data, we should have \(\mathbb{E}[\textbf{s}] = \textbf{0}\) since by an argument of symmetry, a tilt in any direction should be equally likely. Assume now we add some drift \(\textbf{k}\) giving rise to new shear parameters \(\hat{\textbf{s}}\). Looking then at the expectation of \(\mathbf{\hat{s}}\). Since the composition of shear transformations simply amounts to adding the shear parameters, we get
\begin{equation}
\mathbb{E}[\mathbf{\hat{s}}] = \mathbb{E}[\mathbf{s}+\textbf{k}]
= \mathbb{E}[\mathbf{s}]+\mathbb{E}[\textbf{k}]
= \textbf{k} \enspace .
\end{equation}
Thus, given \(N\) fitted ellipsoids with \(\textbf{s}^{(i)}\) the vector of shear constants for ellipsoid \(E_i\), \(1 \leq i \leq N\), we estimate the drift in the images \(\textbf{k}\) simply by the average drift,
\begin{equation}
\textbf{k} = \frac{1}{N} \sum_{i=1}^N \mathbf{\hat{s}}^{(i)} \enspace .
\end{equation}
Enumerating the image sections by \(I_{j}\), \(1 \leq j \leq M\) such that \(I_{1}, \dots, I_{N}\) are ordered with increasing \(z\) choosing \(I_{1}\) as the reference image, drift correction can be obtained by transforming \(I_{j}\) by \(S^{-(j-1)}\).

\subsection*{Drift correction assuming varying drift}

Since drift in images may vary, e.g., due to manual correction during the scanning operation, movement of the sample, or charge equalization, it is likely that the amount of drift varies across sections. Given a large enough population of ellipsoids, it is possible to give an estimate of the drift per image section.\\

Let \(E_i\) denote the \(i\)'th fitted ellipsoid with \(\mathbf{\hat{s}}^{(i)}\) its shear parameters, and let \(1_{E_i \in I_j}\) be an indicator function where \(E_i \in I_j\) is true when \(E_i\) is present in section \(I_j\). Assuming there's an ellipsoid present in every image section \(I_j\), we can define \((\mathbf{k}_j)_1^M\), the sequence of drift parameters estimated per image section given by a discrete function estimate as
\begin{equation}
\textbf{k}_j = \sum_{i=1}^N \frac{1_{E_i \in I_j}}{\sum_{n=1}^N 1_{E_n \in I_j}} \, \mathbf{\hat{s}}^{(i)} \quad , \quad 1 \leq j \leq M \enspace .
\end{equation}
If there exist sections with no ellipsoids, we suggest either interpolating the drift parameters from nearby known values or assume the drift is zero, depending on the dataset. It's worth noting that there's an implicit smoothing present in the above local drift estimation since the ellipsoids are estimated across multiple sections. What we get in return is a more reliable estimate since we enforce the vesicle model on the estimate.



\section*{Experiments}

Because the ground truth drift in FIB-SEM images are unknown, we are from the onset very limited in how well the methods can be validated on such images. We thus look first at synthetic images with artificial known drift focusing both on our ability to estimate the added drift, as well as evaluating the effect parameters choices such as the number of vesicles and the magnitude of the drift. On real images we assess only the perceptual quality of the correction.

\subsection*{Synthetic Data}

\begin{figure}
  \centering
  \includegraphics[trim={0.35cm 0 0.35cm 0},clip,width=\columnwidth]{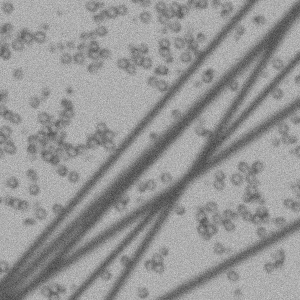}
  \caption{An image section of the synthetic dataset generated for this work.}
  \label{fig:synthetic_image_example}
\end{figure}

To generate synthetic images, we first initialize an image array of $350^3$ voxels in size. We then randomly place vesicles by choosing a random point, generate random ellipsoid radii and rotation parameters to generate the corresponding algebraic matrix as in \ref{eq:ellipsoid_parameter_matrix}. We check the ellipsoid does not overlap with existing ellipsoids before drawing the boundary. An example image can be seen in Fig. \ref{fig:synthetic_image_example}.

\begin{figure}
  \centering
  \includegraphics[trim={0.35cm 0 0.35cm 0},clip,width=\columnwidth]{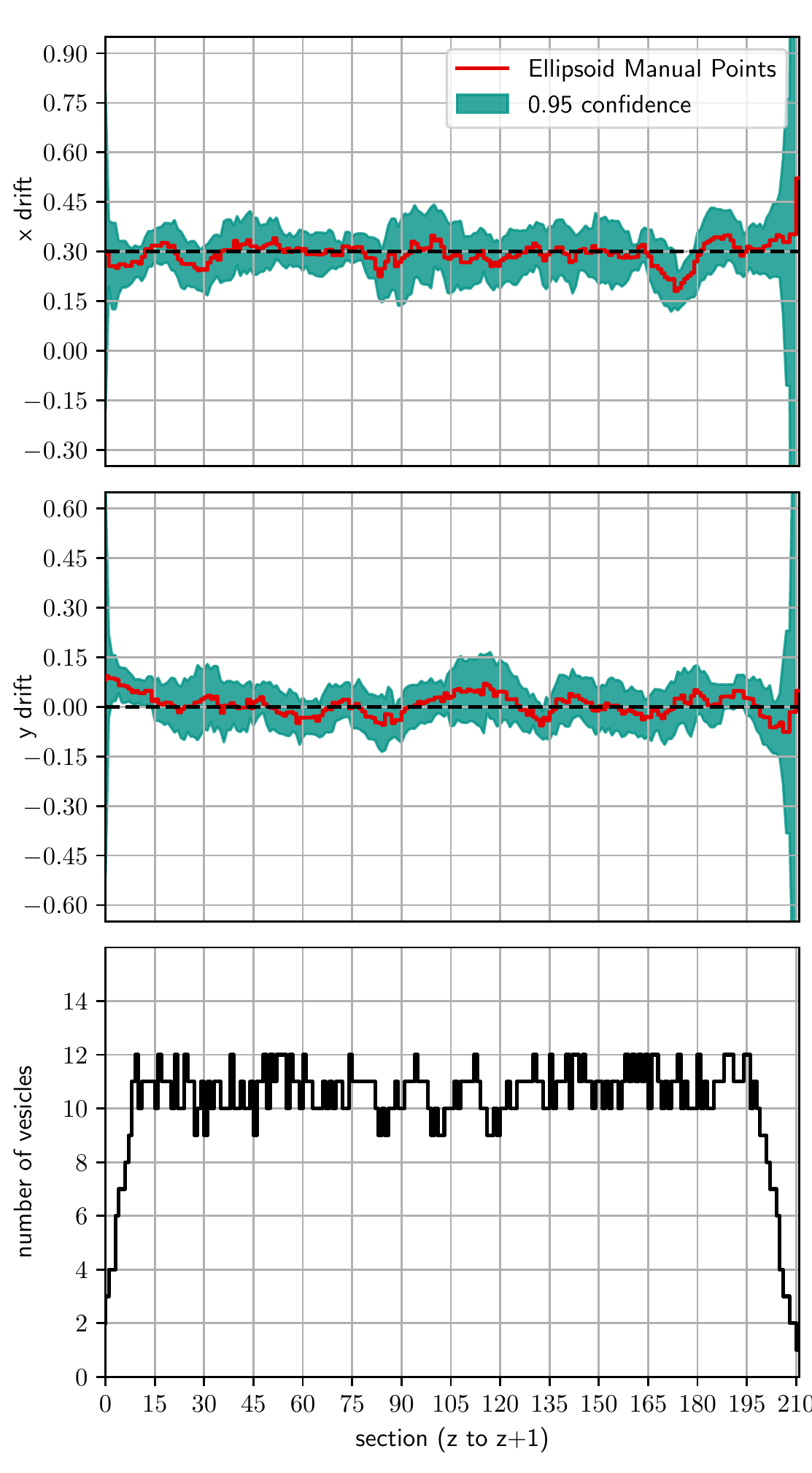}
  \caption{Resulting drift estimate plotted alongside 95\% confidence interval.}
  \label{fig:section_by_section_drift}
\end{figure}

Boundary point annotation was then carried out by hand on the images before fitting ellipsoids and estimating the drift. The per-slice estimated drift can be seen in Fig. \ref{fig:section_by_section_drift}. We notice the confidence of the drift estimate depends expectedly on the number of vesicles used in the estimate.

\subsection*{Real Data}

\begin{figure}
  \centering
  \includegraphics[trim={0.25cm 0 0.2cm 0},clip,width=\columnwidth]{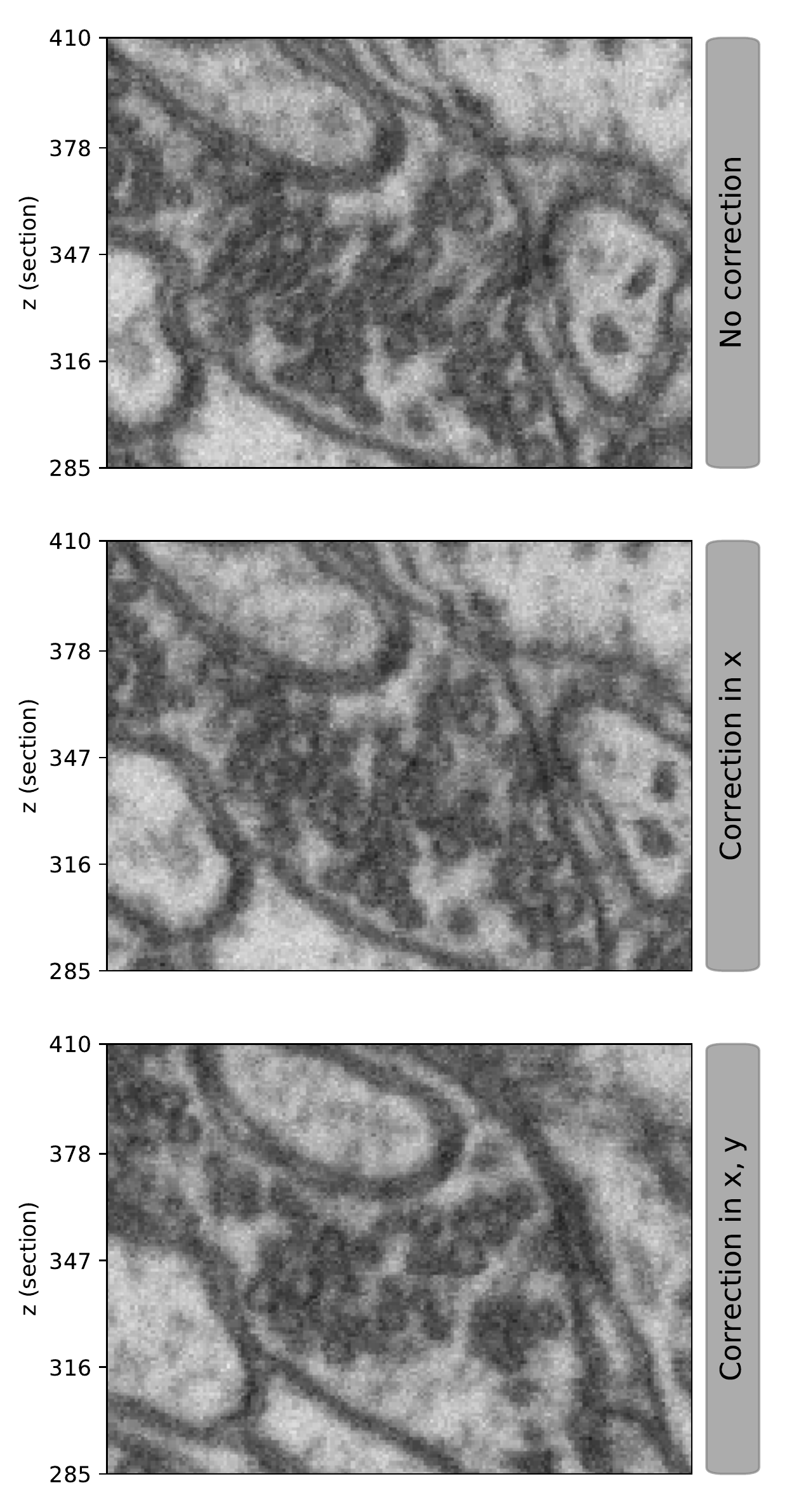}
  \caption{Side view of the dataset showing the correction results on a region with severe drift. As the correction is done in both directions in the image plane, the final image only corresponds to the others at the center row.}
  \label{fig:corrected_images}
\end{figure}

We experiment first on Real FIB-SEM images from the CA1 hippocampus brain region of a healthy adult rodent (see \cite{lausanneDataset} for further details on the dataset). We first manually create a points set of the vesicle cell membrane of a \(~900\) individual vesicles in order to fit ellipsoids directly. We calculate fit the ellipsoid parameters and estimate the drift based. An example area containing substantial drift, can be seen in Fig. \ref{fig:corrected_images} alongside the resulting corrected images.

\section*{Discussion and Conclusion}
In this paper, we present a novel and highly accurate method for correcting drifted FIB-SEM images of neuronal tissue. The method leverages the spherical nature of vesicles and removes the drift by translating the images such that this property is maximized. To complete this task, we use the ellipsoid as a model in order to formulate a theoretical expression of the drift given a family of vesicles. To estimate the vesicles by ellipsoids, fit ellipsoid to pre-segmented boundary points of the vesicles.

Experiments show this method is accurate down to the sub-pixel level with easily acceptable degree of uncertainty, only showing inaccuracy in areas with few vesicles as would be expected. Furthermore, our method outperforms state-of-the-art registration approach that both underestimate the drift, and which are all biased by the presence of synthetically added membranes.

To mitigate the error of smoothing which happens because the ellipsoids are being estimated across multiple image sections, further improvements are possible by doing a more local estimate of the drift. However, in our experience, this causes significantly reduced accuracy. Alternatively, it might be possible to formulate a registration approach that registers the entire image using only information from vesicle regions in isolation.

In our experience, these types of corrections are performed primarily by hand today. Given the sub-pixel magnitude of the drift found in FIB-SEM images, and given that a small sub-pixel drift accumulates to a large discrepancy in distance measures across multiple sections. We believe such manual correction carries significant errors with it. We, therefore, suggest further work should be carried out to assess the effect of the drift on biological images and that more methods are developed for estimating and correcting this drift.